\documentclass[aps,pra,preprint,superscriptaddress, showpacs]{revtex4-1}

\usepackage{mathptmx,MnSymbol,graphicx,color}
\usepackage[normalem]{ulem}
\usepackage{hyperref}

\begin{document}

\title{Electron correlations in the antiproton energy loss distribution in He}
 
\author{S. Borb\'ely}
\email[]{sandor.borbely@phys.ubbcluj.ro}
\affiliation{Faculty of Physics, Babe\c{s}-Bolyai University, 400084 Cluj-Napoca, Romania, EU}

\author{X.-M. Tong}
\affiliation{Center for Computational Sciences, University of Tsukuba, 1-1-1 Tennodai, Tsukuba, Ibaraki 305-8577, Japan}

\author{S. Nagele}
\affiliation{Institute for Theoretical Physics, Vienna University of Technology, 1040 Vienna, Austria, EU}

\author{J. Feist}
\affiliation{Departamento de F\'isica Te\'orica de la Materia Condensada and Condensed Matter Physics Center (IFIMAC), Universidad Aut\'onoma de Madrid, E-28049 Madrid, Spain, EU}

\author{I. B\v rezinov\'a}
\affiliation{Institute for Theoretical Physics, Vienna University of Technology, 1040 Vienna, Austria, EU}

\author{F. Lackner}
\affiliation{Institute for Theoretical Physics, Vienna University of Technology, 1040 Vienna, Austria, EU}

\author{L. Nagy}
\affiliation{Faculty of Physics, Babe\c{s}-Bolyai University, 400084 Cluj-Napoca, Romania, EU}

\author{K. T\H{o}k\'esi}
\affiliation{Institute of Nuclear Research, Hungarian Academy of Sciences (ATOMKI), P.O. Box 51, H-4001 Debrecen, Hungary, EU}
\affiliation{ELI-ALPS,ELI-HU Non-profit Ltd., Dugonics t\'er 13, H-6720 Szeged, Hungary, EU}

\author{J. Burgd\"orfer}
\affiliation{Institute for Theoretical Physics, Vienna University of Technology, 1040 Vienna, Austria, EU}

\date{\today}

\begin{abstract}
We present ab-initio calculations of the electronic differential energy transfer (DET) cross-sections for antiprotons with energies between $3$keV and $1$MeV interacting with helium. By comparison with simulations employing the mean-field description based on the single-active electron approximation we are able to identify electron correlation effects in the stopping and straggling cross sections. Most remarkably, we find that straggling exceeds the celebrated Bohr straggling limit when correlated shake-up processes are included. 
	
\end{abstract}
\maketitle 
\section{Introduction}\label{sec:int}
Inelastic collisions of charged particles with matter probe the response of many-electron systems ranging from linear response in the perturbative limit to the strong-field non-linear response in the non-perturbative regime at low projectile velocities. The characteristic energy loss, stopping power, and energy straggling (the second moment of the energy loss distribution) are among the most important variables quantifying this response. Their investigation dates back to the early work by Bohr \cite{bohr13,bohr15} more than one hundred years ago and continues up to date \cite{bohr48,bethe1930,bailey,bloch1933,landau44,lindhard54,bonderup,andersen1978,besenbacher1980,ahlen80,sigmund96,sigmundbook,sigmund2001}. Present interest in the energy loss distribution is derived from both fundamental aspects of inelastic many-body physics as well as a host of technological and radiation physics applications. The most prominent examples of the latter include hadron-therapy protocols in oncology, sub-surface layer deposition in semi-conductors, and material protection against long-term radiation exposure for space exploration. 

Only recently, progress in methods for exact numerical solutions of the time-dependent many-electron problem and the increased availability of computational power has opened up opportunities for fully ab-initio simulations of the many-electron response to charged particle penetration. The prototypical case in point, for which a - within the numerical accuracy - exact solution is nowadays possible is the inelastic scattering of antiprotons with helium \cite{bailey,borbely14}. This system constitutes the benchmark for the inelastic many-body response and for the energy loss distribution in inelastic collisions for several reasons: helium is the simplest atomic system where correlation effects play a prominent role. Antiprotons are the simplest case of a hadronic projectile that provides a time-dependent Coulomb field driving excitation and ionization without adding complications associated with the charge-transfer channel. Moreover, comparison between proton and antiproton projectile scattering allows for the exploration of the Barkas effect \cite{barkas63}, the variation of the many-electron response under charge conjugation. Pioneering computational studies of correlated two-electron charged particle induced processes in He including the Barkas effect in double ionization~\cite{reading1987a,reading1987b,ford94} and correlation effects in ionization of helium~\cite{reading96,reading97} were performed by Reading and Ford using the forced impulse approximation~\cite{ford1985}. Nowadays, for $\rm{\bar p + He}$ collisions the time-dependent Schr\"odinger equation (TDSE) for the two-electron problem can be solved in its full dimensionality without any approximation.

On the experimental side, the low-energy antiproton ring (LEAR) at CERN has allowed to study fundamental scattering and recombination processes involving antiprotons \cite{sigmund2001,borbely14,barkas63}. The extra-low energy antiproton (ELENA) ring is expected to significantly increase the flux of antiprotons usable in scattering experiments in the near future \cite{schiwietz96}. First full quantum calculations for $\rm{\bar p + He}$ beyond perturbative calculations were performed within the single-active electron (SAE) model by Schiwietz et al.~\cite{schiwietz96} using an atomic-orbital (AO) expansion and by L\"uhr and Saenz \cite{luhr} employing a semiclassical close-coupling approach to the effective one-electron TDSE for $\rm{\bar p + He}$ using a B-spline basis for the radial wave functions. They found sizeable disagreement with the first stopping power measurement by Agnello et al.~\cite{agnello,lodi04} for helium both below and above the stopping power maximum and attributed the discrepancies with the experiment at lower energies to multi-electron or correlation effects neglected within the SAE model. A step towards partially including those were very recently taken by Bailey et al.~\cite{bailey} using a multi-configuration expansion of the He target wave function within the convergent close-coupling (CCC) approach. True two-electron processes such as double ionization and excitation-ionization were, however, still approximated by sequential one-electron excitation and ionization of He and He$^+$.  

For straggling, i.e. the second moment of the energy loss distribution, available experimental data as well theoretical results are still remarkably scarce despite its importance for applications. For gas-phase targets only very few measurements are available \cite{bonderup,andersen1978,besenbacher1980,Vockenhuber}. Theoretical treatments, to date, rely on perturbation theory converging to the high-energy limit $T_B=4\pi Z_p^2Z_Te^2$ for electronic straggling derived by Bohr from classical binary encounter scattering \cite{bohr15} of the projectile on $Z_T$ independent free electrons of the target atom. Remarkably, at non-asymptotic energies ab-initio simulations appear to be still missing up to date.

In the present communication we present first fully ab-initio simulations of the electronic energy loss distribution for antiproton scattering at helium atoms. The two-electron response is treated - within the limits of numerical convergence - exactly and allows, for the first time, to clearly identify the influence of electronic correlations on the energy loss distribution. Most notably, multi-electron shake-up processes yield energy loss fluctuations in excess of the celebrated Bohr straggling limit $T_B$. Atomic units are used unless stated otherwise.

\section{Theoretical methods}\label{sec:theoback}
\subsection{Background} 
The passage of charged particles through matter with atom number density $N$ and thickness $\Delta x$ is accompanied by an energy loss resulting for an initially mono-energetic beam with energy $E_p=\frac{1}{2}m_pv_p^2$ in energy loss distribution $P(\varepsilon)$ with $\varepsilon=E-E_p$, the energy transferred to the target atoms. For dilute matter such as gas targets where non-linear density effects can be safely neglected, $P(\epsilon)$ is related to the differential energy transfer (DET) cross section, $d\sigma(\varepsilon)/d\varepsilon$, as
\begin{equation}\label{eq:P(E)}
P(\varepsilon) = N\Delta x \frac{d\sigma(\varepsilon)}{d\varepsilon}.
\end{equation}
The mean energy loss, the first moment of $P(\varepsilon)$, is, accordingly, given by 
\begin{equation}\label{eq:mean_E}
	\langle \Delta E\rangle = N\Delta x S
\end{equation}
with
\begin{equation}\label{eq:S}
	S=\int \epsilon\frac{d\sigma(\epsilon)}{d\epsilon}d\epsilon
\end{equation}
the energy loss cross section $S$, the mean loss per target atom. The so-called stopping power or stopping force, ($-\frac{dE}{dx}$) follows from Eqs. (\ref{eq:mean_E}) and (\ref{eq:S}) as  
\begin{equation}\label{eq:stopping_p}
	-\frac{\langle \Delta E \rangle}{\Delta x} = N(-S),
\end{equation}
where the minus sign indicates energy lost by the projectile and transferred to the electronic degrees of freedom of the target atom. Likewise, the straggling parameter $\Omega^2$ related to the second moment of the DET follows as
\begin{equation}\label{eq:straggling}
	\Omega^2 = N\Delta xT,
\end{equation}
with
\begin{equation}\label{eq:T}
	T=\int\epsilon^2\frac{d\sigma(\epsilon)}{d\epsilon}d\epsilon 
\end{equation}
referred to as the atomic straggling cross section. $T$ is a measure for fluctuations in the energy loss distribution.

We will focus in the following on the energy transfer to the electronic degrees of freedom. Energy transfer to the He nucleus (''nuclear stopping'') is negligible at high collision energies~\cite{schiwietz96} and provides only a small correction to the stopping cross section of $\leq 10\%$ even at lowest energies ($E_p=3$~keV) considered here. Also for higher moments of the energy loss distribution, the nuclear scattering channel may contribute only a small tail extending to high energies due to rare "hard" binary collisions at a (screened) Coulomb potential. Nuclear contributions can be readily accounted for by elastic binary collisions at a screened Coulomb potential and will be, for completeness,  included when we compare with experiments. We also note that for transmission through dense gas targets, the energy loss distribution $d\sigma(\epsilon)/d\epsilon$ resulting from the individual atomic collisions should be self-convoluted in a multiple scattering setting. Our focus in the following is on single collisions at a multi-electron atom in a dilute gas target.

Early theories on the stopping power ($-\frac{dE}{dx}$) or stopping force based on either classical binary collision approximations \cite{bohr13,bohr15} or first-order quantum approximations \cite{bethe1930,bloch1933} can be written in terms of the  dimensionless so-called stopping number $L(E)$ as 
\begin{equation}
	\label{eq:dedx}
	-\frac{dE}{dx} = N \frac{4\pi e^2Z_p^2Z_T^2}{m_ev_p^2} L(E),	
\end{equation}	
with $Z_p$ ($Z_T$) the nuclear charge of the projectile (target), $v_p$ the speed of the incident projectile, $m_e$ the mass of the electron and $N$ the number density of the target atoms. Well-known approximations to the stopping number include the classical Bohr logarithm
\begin{align}\label{eq:L_Bohr}
	L_{\rm Bohr}(E) = \ln{\frac{1.123 m_ev_p^3}{Z_pe^2\omega}},
\end{align}	
with $\omega$ the classical oscillator (or mean transition) frequency and the Bethe logarithm derived from the first Born approximation
\begin{align}\label{eq:L_Bethe}
	L_{\rm Bethe}(E) = \ln{\frac{2 m_ev_p^2}{\hbar\omega}}.
\end{align}	
A multitude of more sophisticated approximations have been developed over the years approximately including
corrections for the Barkas effect, binding shell corrections, so-called "bunching" effects accounting for deviations from the independent-electron response in atoms and solids as well as interpolations between the low-energy regime and the high-energy regime where the Bohr approximation applies covering the stopping maximum \cite{sigmundbook,sigmund2001}.
\subsection{Semiclassical impact-parameter approach}
The numerical solution of the time-dependent Schr\"odinger equation (TDSE) for a non-perturbative treatment of the electronic DET cross section involves, generally the (semiclassical) impact parameter (IP) approach. Accordingly, the projectile is treated as a classical charged particle moving on a straight-line trajectory $\vec{R}(t)=\vec{b}+\vec{v}_pt$. Here $\vec{b}$ is the impact parameter vector, and $\vec{v_p}$ is the projectile's velocity. In turn, the electronic dynamics driven by the time-dependent Hamiltonian $H(t)$ is treated fully quantum mechanically by solving the TDSE. The IP approximation is well justified and leads to negligible errors for the antiproton energies above a few keV, the projectile energies considered in the following. The impact parameter dependent transfer probability density $P_{i\rightarrow f}(\varepsilon;b,v_p)$ from the initial state $i$ to the final state $f$ representing excitation or ionization is determined by the projection of the numerically evolved state at time $t_t$ $\left |\Psi(b,v_p,t_t)\right\rangle$, parametrically dependent on impact parameter and projectile velocity, onto the corresponding exit-channel state $|\psi_f(E_f)\rangle$,
\begin{equation}\label{eq:prob}
 P_{i\rightarrow f}(\epsilon; b,v_p) =|\langle\psi_f(E_f)|\Psi(b,v_p,t_t)\rangle|^2,
\end{equation}
where $\epsilon=E_f-E_i$ with $E_i$ the energy of the initial state (i.e.~the ground state of the target), and $E_f$ the energy of the final state (excited, singly, and doubly ionized states) at the termination point $t_t$ of the time propagation. As $t_t$ is finite in a realistic numerical simulation, $P_{i\rightarrow f}$ must be tested for convergence as a function of $t_t$. In Eq. (\ref{eq:prob}), the rotational symmetry of the He initial state was used as $P_{i\rightarrow f}$ depends only on the magnitude $b$ of the impact parameter vector. From Eq. (\ref{eq:prob}) the differential energy transfer cross section follows as
\begin{equation}
 \frac{d\sigma}{d\varepsilon}(\varepsilon)=2\pi\int db b\sum\limits_{f} P_{i\rightarrow f}(\varepsilon; b, v_p),
 \label{eq:det}
\end{equation}
where the sum extends over those degenerate final states $f$ that contribute to the fixed energy transfer $\varepsilon$.

The total energy loss or stopping cross section can be expressed as
\begin{equation}
S(v_p)=2\pi\int b S(b;v_p) db, 
\label{eq:sint}
\end{equation}
where $S(b;v_p)$ is the impact parameter dependent mean energy loss given in terms of the loss distribution (Eq.~\ref{eq:prob}) by
\begin{equation}
S(b;v_p)=\sumint_f \varepsilon P_{i\rightarrow f}(\varepsilon,b;v_p)  d\varepsilon
\label{eq:lossindirect}
\end{equation}
Analogously, the straggling cross section reads
\begin{equation}
T(v_p)=2\pi\int bT(b;v_p)db,
\label{eq:tint}
\end{equation}
where $T(b;v_p)$ is the impact parameter dependent straggling which can be calculated from the energy transfer probability density
\begin{equation}
T(b;v_p)= \sumint_f P_{i\rightarrow f}(\varepsilon,b;v_p) \left[\varepsilon - S(b;v_p)\right]^2d\varepsilon.
\label{eq:straggindirect}
\end{equation}

Alternatively to the explicitly channel-resolved expressions Eq. (\ref{eq:lossindirect}) and Eq. (\ref{eq:straggindirect}), $S$ and $T$ can be directly expressed in terms of expectation values of the unperturbed electronic Hamiltonian $H_0$ calculated with the initial ($E_0$) as well as the evolved state $\left |\Psi(b,v_p,t) \right\rangle$,
\begin{equation}
 S(b;v_p)=\left\langle E \right\rangle -E_0
 \label{eq:stopping_direct}
\end{equation}
with
\begin{equation}
 \left\langle E \right\rangle = \left\langle \Psi(b,v_p,t)  | H_0 | \Psi(b,v_p,t) \right\rangle,
\label{eq:enexp}
 \end{equation}
and
\begin{equation}
 T(b;v_p)= \left\langle E^2 \right\rangle -2\left\langle E \right\rangle\left[E_0+S(b;v_p)\right]+\left[E_0+S(b;v_p)\right]^2
 \label{eq:straggling_direct}
\end{equation}
with
\begin{equation}
 \left\langle E^2 \right\rangle = \left\langle \Psi(b,v_p,t)  | H_0^2 | \Psi(b,v_p,t) \right\rangle.
 \label{eq:en2exp}
\end{equation}
Within a fully converged calculation and in the limit $t\rightarrow \infty$, Eqs. (\ref{eq:stopping_direct},\ref{eq:straggling_direct}) would be equivalent to Eqs. (\ref{eq:lossindirect},\ref{eq:straggindirect}). However, since the numerical propagation must be terminated at a finite time $t_t$ when both the departing antiproton and the ionized electron are still at a moderately large distance from the He target as well as from each other, the projections Eq. (\ref{eq:prob}) as well as the expectation values Eqs. (\ref{eq:enexp},\ref{eq:en2exp}) may be affected by, in general different, termination errors. We estimate the size of such errors by comparing $S$ and $T$ calculated by the two alternative methods. 
\subsection{Time-dependent close coupling method}
For accurate energy loss values and energy loss distributions a high-precision description of the collision between the projectile and one target atom is required. In order to achieve this goal, we numerically solve the time-dependent Schr\"odinger equation describing the quantum dynamics of the two active electrons of the He target in the presence of the passing-by antiproton \cite{borbely14}.

The time-dependent Hamiltonian is given by
\begin{equation}
 H(t) = H_0 +\sum\limits_{i=1}^2\frac{1}{|\vec r_i-\vec R(t)|},
 \label{eq:ham}
\end{equation}
with $H_0$ the unperturbed electronic Hamiltonian of the helium atom
\begin{equation}
 H_0=\sum\limits_{i=1}^2\left ( -\frac{\nabla_i^2}{2} -\frac{2}{r_i} \right)+\frac{1}{|\vec r_1-\vec r_2|}.
\label{eq:ham0}
 \end{equation}
We solve the TDSE
\begin{equation}
 i\frac{\partial \Psi(t)}{\partial t}=H(t)\Psi(t)
\end{equation}
by the time-dependent close-coupling (TDCC) method~\cite{borbely14,foster08,feist08}. Briefly, the fully correlated two-electron wave function is represented in the basis of symmetrized coupled spherical harmonics~\cite{borbely14}, while the radial partial wave functions are represented using the finite element discrete variable representation (FEDVR) method \cite{schneider05,rescigno00}, where each radial coordinate is divided into segments with variable length (i.e. finite elements - FEs). Then, inside each FE the radial wave function is represented on a local polynomial basis (i.e. discrete variable representation - DVR) built on top of a Gauss-Lobatto quadrature to ensure the continuity at the FE boundaries.

For the temporal propagation of the wave function the short iterative Lanczos (SIL) method with adaptive time steps is applied \cite{park86,schneider2011}.  The time-evolution of our system is started with the projectile located at $R_z=-40$~a.u., and with the ground state He target located at the center of our coordinate system.  The ground state of helium was obtained by propagating an initial trial wave function in negative imaginary time ($t \rightarrow -i\tau$). The time-propagation is continued up to the termination time $t_t$ at which the position $R_z=80$~a.u. (the distances from the He atom at zero impact parameter) of the antiproton is reached. For channel-resolved energy transfer densities [Eq.(\ref{eq:prob})] we project onto asymptotic channel wave functions, which are constructed as a symmetrized product of single-electron wave functions, thus they neglect the electron-electron and electron-projectile interactions in the continuum. These are correct only in the limit $R\rightarrow \infty$ and, if ionization is involved, $r_i\rightarrow\infty$. Therefore, errors due to the finite propagation time needs to be checked.
\subsection{Mean-field approximation}
\label{sec:MFA}
In order to quantify the role of correlations in the DET distribution and to compare with previous non-perturbative calculations for stopping \cite{bailey,schiwietz96,cabrera05,luhr} we perform in parallel mean-field simulations. For the calculation of the DET distribution, they involve two separate approximations to be kept track of. The first one is the approximation of the exact Hamiltonian by the sum of two effective single-electron Hamiltonians
\begin{equation}
 H(t)=\sum\limits_{j=1}^2H_{j}^\mathrm{eff}(t),
\end{equation}
with
\begin{equation}
 H_{j}^\mathrm{eff}(t)=-\frac{\nabla_j^2}{2}+V_\mathrm{eff}(r_j)+\frac{1}{|\vec r_j-\vec R(t)|}
\end{equation}
where the effective mean-field potential $V_\mathrm{eff}$ accounts for the nuclear Coulomb field and the mean screening field provided by the other electron. Using a static screening potential as in the following,
\begin{equation}
 V_\mathrm{eff}(r)=-\frac{Z_c+a_1e^{-a_2r}+a_3re^{-a_4r}+a_5e^{-a_6r}}{r},
 \label{eq:modpot}
\end{equation}
where $Z_c$ is the charge of the residual ion and the model parameters are taken from \cite{tong05} in the TDSE
\begin{equation}
 i\frac{\partial \Psi_j(t)}{\partial t}=H_j^\mathrm{eff}(t)\Psi_j(t)
 \label{eq:tdseSAE}
\end{equation}
leads to the single-active electron (SAE) approximation~\cite{tong01,tong00,yao93,tong02}.

Alternatively, within time-dependent density functional theory (TDDFT), $V_{\mathrm{eff}}$ contains dynamical screening due to the self-consistent coupling of the evolution to the time-dependent electronic density $\rho = \sum\limits_j|\Psi_j|^2$ \cite{gross84,hohenberg64,bauer97,tong98}. Within the TDDFT approach, correlation effects can be taken into account on the mean-field level.

Final state probabilities for excitation (EX) and ionization (I) follow from the projection amplitudes $P_f^{(1)}=\left|\langle \Psi_f| \Psi_j \rangle\right|^2$. Unlike for the projection of the fully correlated two-electron wave function, these $P_f^{(1)}$ are one-electron probabilities on a mean-field level. Therefore, to account for multi-electron processes (specifically in the case of He, two-electron processes), a second approximation is invoked, the independent event model (IEM). This applies to both the SAE and TDDFT approaches. Accordingly the joint probability for ionizing, e.g., one and exciting the other electron is approximated by $P_{EX-I}=2P_{EX}^{(1)}P_{I}^{(1)}$.
Analogously, double ionization (DI) is approximated by $P_{DI}=P^{(1)}_I P^{(1)}_I$. Such an IEM for multi-electron processes can be modified to account for an assumed sequentiality of these processes. Eg. sequential double ionization is expressed as $P_{DI}^{\mathrm{Seq.}}={P_{I}^{(1)}}^+P_{I}^{(1)}$, where $P_{I}^{(1)}$ is the one-electron ionization probability for neutral helium, while ${P_{I}^{(1)}}^+$ is the ionization probability of $\mathrm{He}^+$ calculated from Eq. (\ref{eq:tdseSAE}) with the effective potential [Eq. (\ref{eq:modpot})] reduced to the bare Coulomb potential ($-2/r$).
\subsection{The Bohr model}
The pioneering study of the energy transfer process between an incident charged particle and atomic targets performed by Bohr \cite{bohr13,bohr15} dates back more than a century predating even his quantum atomic model. Apart from historic interest, it still serves as a useful guide for the processes underlying contemporary models for stopping and straggling. Bohr's model for energy loss involves two contributions: the close collision regime for small impact parameters $b<b_0$ approximated by binary Coulomb scattering between the classical electron on the incident charged particle and the distant-collision regime for large impact parameters for which the projectile supplies the time-dependent electric field that excites the electron with oscillator (i.e. transition) frequency $\omega$ (reminiscent of Thomson's atom model of harmonically bound electrons).

A smooth transition between the two regimes is expected at an intermediate impact parameter $b_0$ which must simultaneously fulfill two requirements: $b_0$ should be large compared to the so-called collision diameter $b_c=Z_pe^2/(m_ev_p^2)$ \cite{bohr13,ahlen80} but small compared to the characteristic impact parameter for resonant excitation of the target electrons $b_r\approx v_p/\omega$ \cite{bohr13,ahlen80} by the time dependent Coulomb field of the passing-by projectile. These two requirements can be combined ($b_c<b_r $) to the criterion for the validity of the Bohr model 
\begin{equation}
v_p > \left(\frac{2Z_pe^2\omega}{m_e}\right)^{1/3}, 
\label{eq:bohrlimit}
\end{equation}
where the classical oscillator frequency $\omega$ should be replaced by a typical quantum excitation frequency the order of magnitude of which is given by the first ionization potential $I_{p_1}$, $\hbar\omega\simeq I_{p_1}$.

Within the framework of this classical model, Bohr also derived the (non-relativistic) high-energy limit for the straggling cross section, i.e. the second moment of the DET which is given by the projectile-energy independent constant
\begin{equation}\label{eq:TB}
	T_B = 4\pi Z_pZ_Te^2.
\end{equation}
For later reference we emphasize that Eq. (\ref{eq:TB}) describes the response of $Z_T$ independent (classical) electrons implicitly invoking the IEM. Remarkably, to this date Eq. (\ref{eq:TB}) has remained the benchmark with which current experimental and theoretical results for straggling are to be compared.

\section{Angular momentum basis convergence}
Since both the TDCC for solving the full two-electron dynamics as well as mean-field models such as the SAE are based on the numerical solution of the time-dependent Schr\"odinger equation, rigorous numerical checks are required. While convergence with respect to size and density of the radial grid, time-propagation parameters, or length of the projectile trajectory has been tested previously \cite{borbely14,tong01}, a critical issue of particular relevance for the present study is the convergence with respect to the number of partial waves or angular momenta included. During energetic binary collisions there is a large energy and momentum transfer from the projectile to the electron, which also implies a large angular momentum transfer. If the truncated angular momentum basis is not large enough to accommodate such angular momentum transfers, then the probability for generating high energy continuum electrons will be significantly suppressed. The importance of including high-angular momentum partial waves is not specific to the numerical solution of the TDSE or to the energy loss but has  been previously observed in a first Born approximation calculation of the angular distribution of high-energy electrons emitted in $\mathrm{p+He}$ collisions~\cite{madison73,manson75}. Since this effect is more pronounced at high projectile velocities, we have performed the angular momentum basis convergence tests at 1 MeV antiproton energy, the highest energy considered in this work.

\begin{figure}
\includegraphics{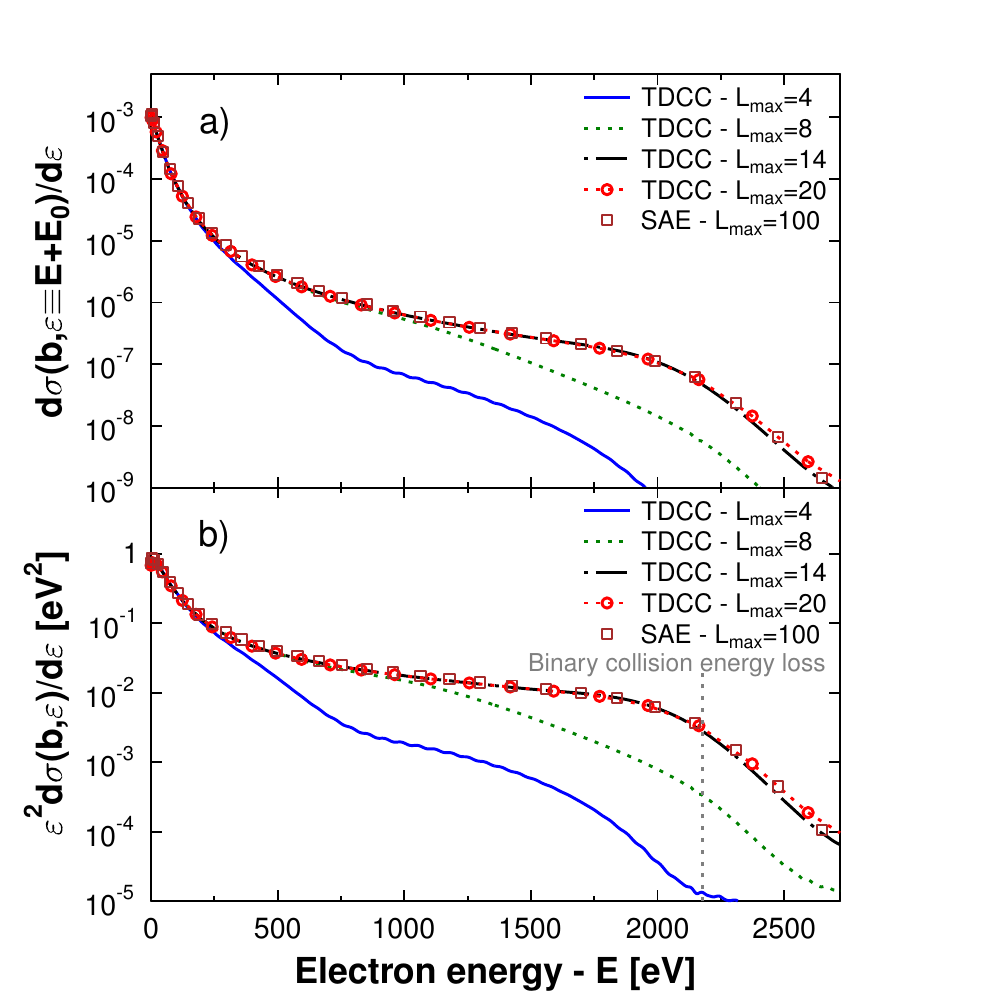}
\caption{ The single ionization (SI0) impact parameter resolved DET $d\sigma(b,\varepsilon)/d\varepsilon$ (a) and the energy transfer square ($\varepsilon^2$) rescaled DET (b) as a function of electron ejection energy for the 1~MeV antiproton projectile at fixed $b=1$~a.u. impact parameter. This impact parameter value was chosen to coincide with the maximum of the impact parameter dependent DET. TDCC and SAE results with different angular momentum basis size ($L_{max}$) are compared for the SI0 single ionization channel . In (b) the binary collision energy loss ($2v_p^2$) is also indicated with a vertical dotted line. \label{fig:angdep}}
\end{figure}

We compare the impact parameter resolved DET, $d\sigma (b,\varepsilon)/d\varepsilon\equiv \sum\limits_{f} P_{i\rightarrow f}(\epsilon; b) $, from the TDCC for the single ionization channel with the second electron remaining in the ground state denoted in the following by SI0 with the corresponding SAE results (Fig. \ref{fig:angdep}a). 
We also checked for $\varepsilon^2 d\sigma (b,\varepsilon)/d\varepsilon$, the DET weighted with the squared energy transfer which places enhanced weight on large energy and angular momentum transfer entering straggling (Fig.~\ref{fig:angdep}b). Obviously, convergence is reached when the  maximum classically allowed binary encounter momentum transfer $\Delta p\sim 2 v_p\simeq 13$~a.u.~at an impact parameter of the order of the atomic radius $b\simeq 1$~a.u.~corresponding to an angular momentum transfer of $L_{max}\simeq b\Delta p\simeq 13$~a.u.~can be accurately represented. 
\begin{figure}
\includegraphics{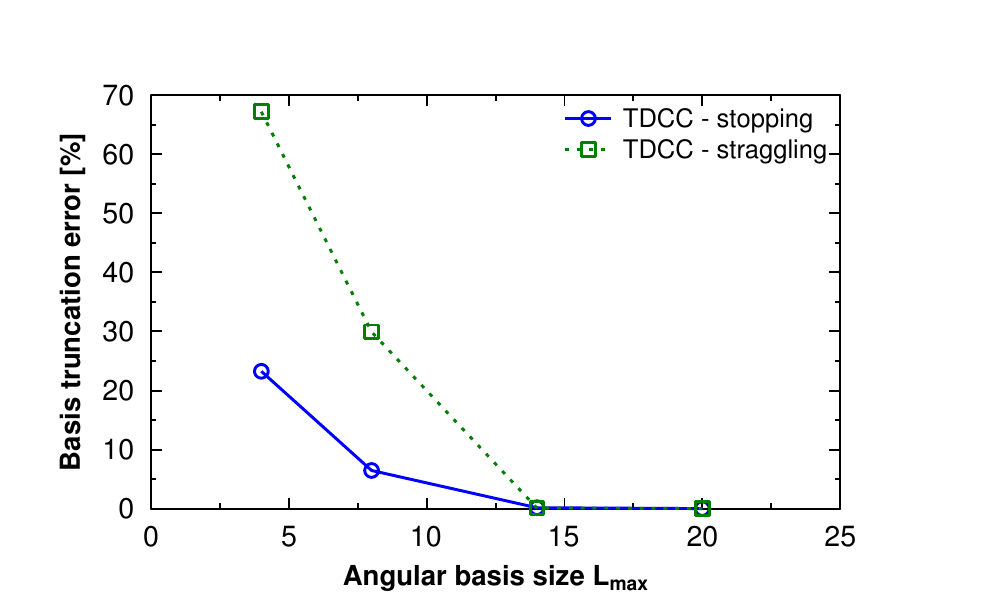}
\caption{The truncation error for stopping and straggling as a function of the angular momentum basis size $L_{max}$ at $E=1$~MeV ($v_p=6.32$~a.u.) and $b=1$~a.u.  \label{fig:angconv}}
\end{figure}
In the case of the full TDCC simulation we choose a highly asymmetric partial wave basis with $0\le l_1\le L_{max}\le 20$ for the ionized electron while $l_2$ is constrained to low angular momentum $l_2=0,1,\dots,l_{2,max}$, where we find convergence already for $l_{2,max}=1$.

The truncation error as a function of the maximum of total (coupled) angular momentum $L_{max}$ included (Fig.\ref{fig:angconv}) shows that previously used small angular momentum basis sizes ($L_{max}\le 6$) for calculation of ionization cross section \cite{borbely14,bailey} are insufficient to accurately account for the stopping and straggling at high energies. We estimate the truncation error by comparison with the reference calculations $S_{ref}$ and $T_{ref}$ in which the contributions of very high $L>>L_{max}\simeq 15$ taken from corresponding SAE calculations are included, as for asymptotically high $L$ the influence of correlation effects can be safely excluded. 

\section{Differential energy loss distributions}
\subsection{Differential energy transfer}
The differential energy transfer (DET) cross section, $d\sigma (\varepsilon) / d\varepsilon$, integrated over the impact parameter [Eq.(\ref{eq:det})] is the key input quantity of interest determining the stopping and straggling.

\begin{figure}
\includegraphics{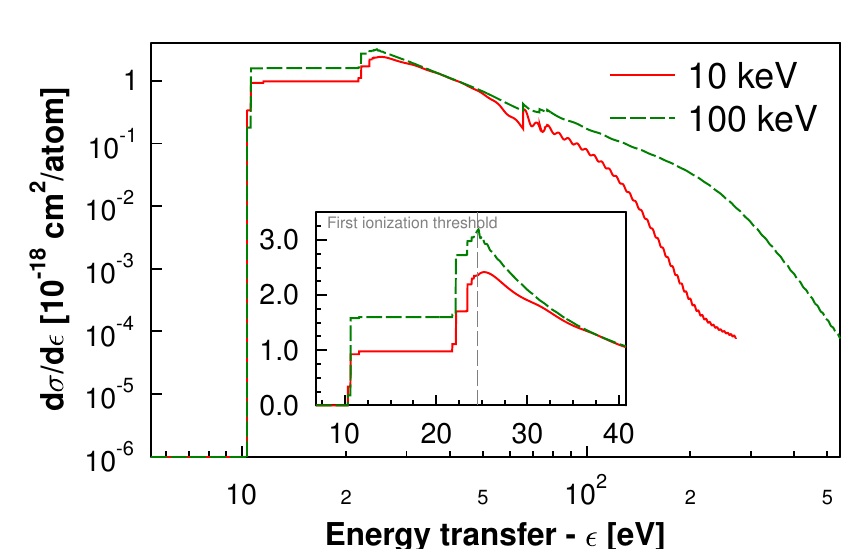}%
\caption{The  energy transfer cross section $d\sigma(\varepsilon)/d\varepsilon $ as a function of $\varepsilon$ for different antiproton impact energies $E$. The inset shows the  finite cross section at and the continuous transition across the first ionization threshold.  \label{fig:etcs}}
\end{figure}

While $d\sigma(\varepsilon)/d\epsilon$ is a continuous function above the first ionization threshold, $I_{p_1}=24.6$~eV, it is discrete below $I_{p_1}$. In order to display the continuity across the threshold we analytically continue $d\sigma (\varepsilon)/d\varepsilon$ for $0\le\varepsilon\le I_{p_1}$ as 
\begin{align}
	\frac{d\sigma}{d\epsilon} = \sum\limits_{n,l,m}\sigma_{nlm}D(n,l)
	\approx\sum\limits_{n,l,m}\sigma_{nlm}(E_{n+1,l}-E_{n,l})^{-1}
\end{align}
with $D(n,l)$ the spectral density of bound states of a given $n,l$ and $E_{n,l}$ the energy of the excited bound state. Both above and below the threshold the multiple (quasi) degeneracies are included. As expected for Coulomb interactions, $d\sigma(\varepsilon)/d\varepsilon$ is continuous and finite across the first ionization threshold (see inset of Fig. \ref{fig:etcs}). At all collision energies, ionization dominates over (exclusive) bound-state excitations (Fig. \ref{fig:etcs}). 
This implies that a typical ''mean'' energy transfer, i.e. the mean value of this distribution, is somewhat larger than $I_{p_1}$ suggesting also a value suitable for $\omega$ in Bohr's model [Eq. (\ref{eq:bohrlimit})]. We observe a power-law behavior of the high-energy tail of $d\sigma (\varepsilon)/d\varepsilon\sim\varepsilon^{-\alpha}$ ($\alpha\simeq 2.2$ for energies below the binary encounter limit) as expected. The DET also contains some fluctuations (most visibly for 10~keV), which are the trace of the unresolved Fano resonances\cite{Kaldun738,Gruson734}. The discontinuities of the DET in the $0.9~\mathrm{a.u.}<\varepsilon < 2.9$~a.u.~energy transfer interval signify the appearance of the ionization-excitation channels. As expected, for each collision energy the high energy tail of the transfer distribution extends to the binary encounter limit $\varepsilon=(2v_p)^2/2$ above which electron emission is strongly suppressed and not resolved in our simulation.
\subsection{Comparison with the Bohr model}
\begin{figure*}
\includegraphics{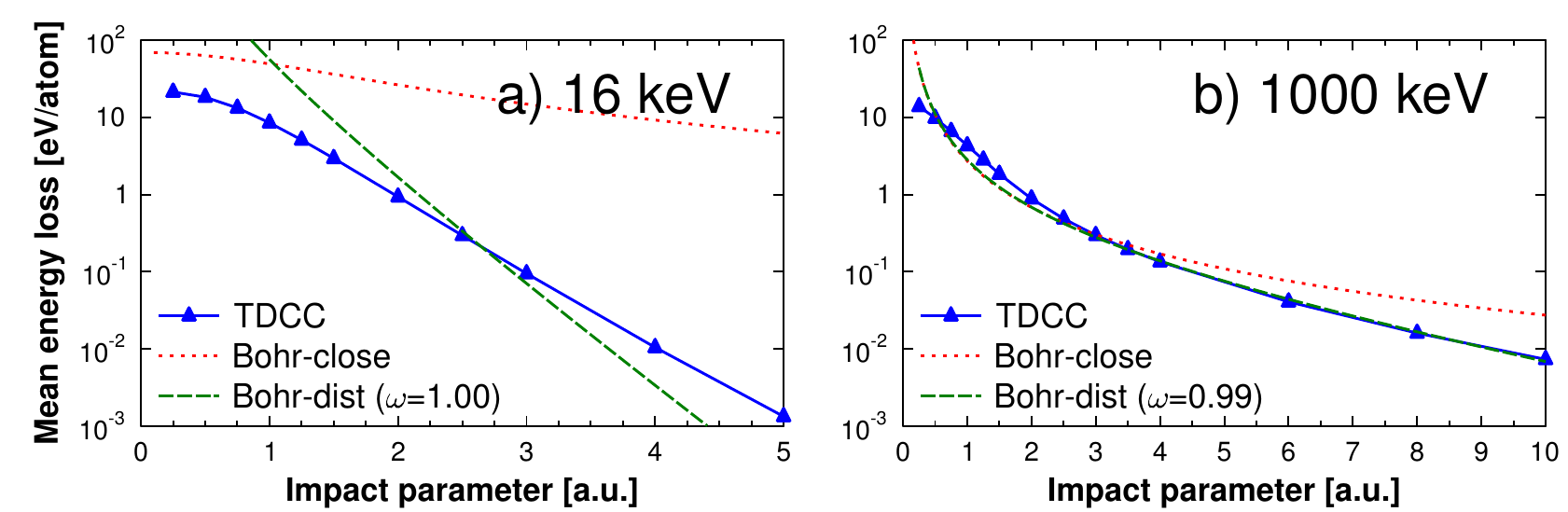}
\caption{The mean energy loss $S(b;v_p)$ [Eq. (\ref{eq:lossindirect})] as a function of the impact parameter for different antiproton energies: 16 keV (a) and 1000keV (b). The present TDCC results contain only the inelastic one-electron channels: single ionization (SI0) and single excitation (EX0) with the second electron remaining in the ground state and are compared to the predictions of the Bohr model for close and distant collisions~\cite{bohr13,ahlen80}. The oscillator frequencies as given in the figures are obtained by fitting the distant collision model to the TDCC results at large impact parameters.\label{fig:bohr}}
\end{figure*}

Another energy loss distribution, differential in impact parameter, but integrated over all energy transfers $S(b;v_p)$ is of considerable conceptual interest. This distribution allows a direct comparison of the TDCC simulation with the original Bohr model for energy loss (Fig.~\ref{fig:bohr}).
Since (correlated) multi-electron processes are not included in the Bohr model we restrict for this comparison the TDCC energy loss to one-electron processes by projecting the evolved state exclusively onto one-electron inelastic channels, i.e pure single ionization (SI0) and pure single excitation with the second electron remaining in the ground state (EX0) [Eq.(\ref{eq:prob})]. It should be noted that in Bohr's close collision model the impact parameter of projectile refers to the quasi-free electron while in the quantum calculation it denotes the distance to the ionic core of the target. Only upon averaging over the ensemble of classical electrons representing the initial bound state the two agree. Moreover, in the close collision model the classical electron is assumed to be at rest in the target frame thereby neglecting the initial momentum-space distribution, i.e. the Compton profile of the initial state.

While at lower projectile energies ($v_p<$1, Fig.~\ref{fig:bohr}a) neither the close-collision contribution expected to be applicable for $b<b_0$ nor the distant-collision contribution for $b>b_0$ approximates the TDCC results well, in the perturbative regime ($E=1$~MeV, $v_p=6.32$, Fig.~\ref{fig:bohr}b) the distant-collision, overall, yields reasonable agreement. The latter is, obviously, related to the fact that, to some extent, it successfully mimics the dipole transitions by virtual photon absorption 
\cite{[{}][{ p. 414}]heitler,[{}][{ ch. 15}]jackson} 
closely related to first-order quantum perturbation theory. For the comparison we have fitted the oscillator frequency $\omega$ in Bohr's distant-collision model to TDCC results for large impact parameters ($b>b_0=v_p/\omega$). The resulting $\omega$ (Fig.~\ref{fig:bohr}) closely matches the expectation of a mean excitation energy slightly above $I_{p_1}=0.9$~a.u. (24.6 eV) as suggested by the DET cross section (Fig.~\ref{fig:etcs}).
As expected, for low antiproton energies ($E_p\sim 16$~keV, Fig.~\ref{fig:bohr}a) and outside the validity of the Bohr model (Eq.~\ref{eq:bohrlimit}), the transition between the close and distant collision regimes is not smooth. With increasing antiproton energies this transition smoothens and at high antiproton energies ($1000$~keV) there is a large impact parameter region where the close and distant collision energy loss predictions overlap. In view of the simplicity of the Bohr model, the agreement between Bohr's distant collision model (with fitted $\omega$) and the high precision TDCC calculations for energies above $\approx100$~keV, when restricted to one-electron processes, is remarkably good. The discrepancies to the close collision model are generally larger, in part due to the neglect of the atomic Compton profile. The latter deficiency can be corrected within the framework of more advanced classical models, in particular the classical-trajectory Monte Carlo method~\cite{abrines66,percival76,reinhold93}.

\subsection{Multi-electron energy loss channels}
By comparing the present TDCC \emph{ab initio} approach with the IEM using SAE calculations (for details  and definitions see Section \ref{sec:MFA}) as input, the importance of different single- and multi-electron energy loss channels and the influence of correlations in each of these can be assessed. To this end we group the final states of the energy-transfer probabilities $P_{i\rightarrow f}(\varepsilon,b,v_p)$ into four different exit channels: single ionization with the second electron remaining in the ground state (SI0), single excitation with the second electron in the ground state as well (EX0), simultaneous single ionization and shake-up excitation of the second electron (SI-EX) and double ionization (DI). We note that the contributions from double excitations leading to formation of autoionizing resonances are implicitly included in the SI0 and SI-EX channels as we do not explicitly project onto them. We note that their contribution to total stopping and straggling is, in particular, at high collision energies negligibly small. The one-electron channels SI0 and EX0 allow for a direct comparison between the TDCC and mean-field models such as the present or previously employed SAE approximations and for probing for electron correlation effects in one-electron transitions. These are to be distinguished from true multi-electron transitions (SI-EX and DI) for which  models \cite{bailey,schiwietz96,luhr} based on SAE approximations, TDDFT or convergent close coupling calculations have been invoked, in addition to, the independent event model (IEM) thereby neglecting explicitly correlated transitions. Such dynamical correlations are fully accounted for by the present TDCC simulation.
\begin{figure*}
\includegraphics{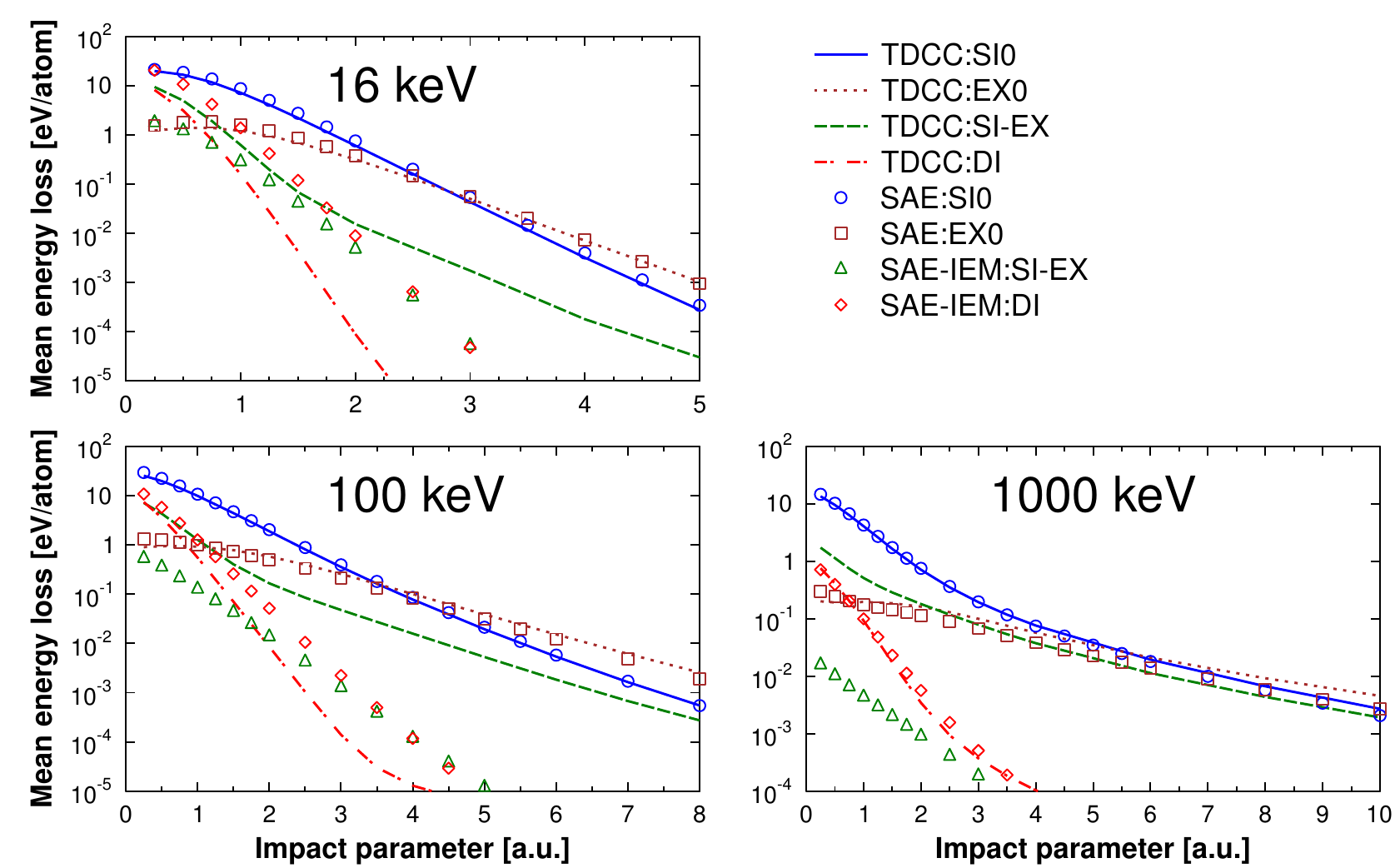}
\caption{Exit-channel decomposition of the energy loss $S(b;v_p)$ as a function of the impact parameter for different antiproton energies (16, 100, 1000 keV) obtained within the framework of the present TDCC and SAE-IEM models. Single ionization (SI0), single excitation (EX0), correlated excitation-ionization (SI-EX), and double ionization (DI). In the SAE approximation the IEM is invoked to approximate multi-electron transitions.\label{fig:Svsb}}
\end{figure*}
Overall, the relative importance of different loss channels varies only weakly over a wide range of collision energies (Fig.~\ref{fig:Svsb}). The one-electron channels dominate, SI0 at small impact parameters and EX0 at large impact parameters. This explains the success of mean-field models for stopping. However, correlated multi-electron process, in particular the SI-EX process provide a significant contribution throughout and become nearly as large as SI0 at large impact parameters and collision energies. It is these processes for which the SAE and similar mean-field models with their uncorrelated IEM extension completely fail (Fig.~\ref{fig:Svsb}).

The SAE-IEM does not account for the correlated ''shake-up'' of the second electron during the ionization process. The importance of such shake-up has recently been also demonstrated in the timing of photoionization by attosecond pulses 
\cite{feist14,Isinger893,ossiander16}.
Also for DI, the SAE-IEM mostly fails, however with the remarkable exception in the perturbative regime at high collision energies. Here the SAE-IEM reproduces the TDCC quite well indicating that direct uncorrelated double ionization dominates over shake-off. 

By contrast, for true one-electron transitions the present SAE yields excellent agreement for SI0 at all energies and impact parameters while for EX0 the agreement is still good with minor deviations observable. The latter can be easily explained by the fact that the final excited state in neutral helium carries the signatures of electron correlations and screening missing in the SAE model.

\begin{figure*}
\includegraphics{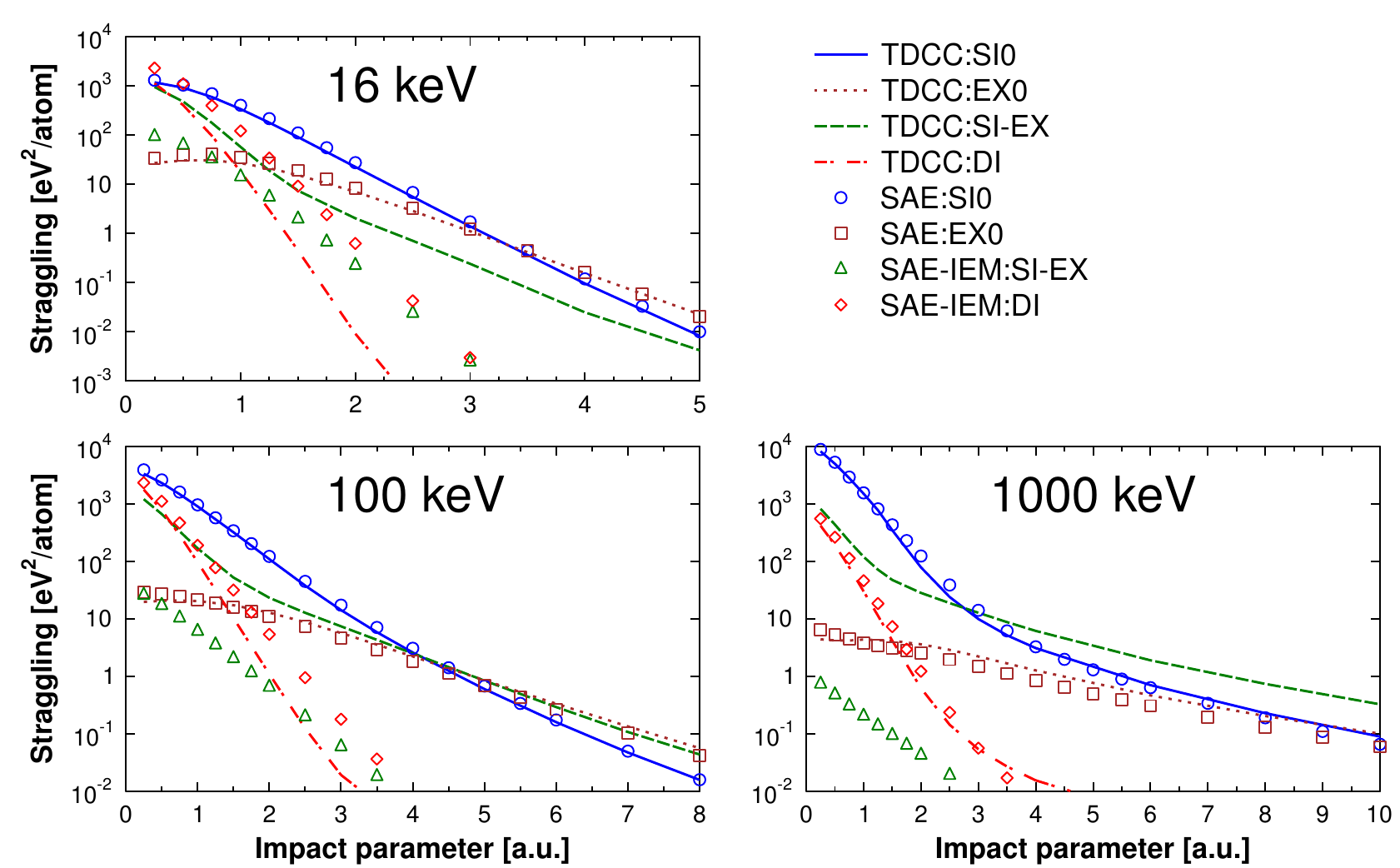}%
\caption{ Exit channel decomposition of the straggling probability $T(b;v_p)$ as a function of the impact parameter for different antiproton energies (16, 100, 1000 keV) obtained in the framework of the present TDCC and SAE-IEM models. Single ionization (SI0), single excitation (EX0), excitation-ionization (SI-EX) and double ionization (DI). Within the SAE approximation, the IEM is used to approximate multi-electron transitions.\label{fig:Tvsb}}
\end{figure*}

The relative importance of the different loss channels changes when higher moments of the energy loss distribution are considered.  Specifically, for the impact-parameter dependent straggling $T(b,v_p)$ (Fig. \ref{fig:Tvsb}) the SI0 channel still provides the largest contribution at small impact parameters, while at large impact parameter values the contributions from the SI-EX and EX0 channels dominate. Most notably, the contribution from shake-up ionization (SI-EX) to energy loss fluctuations is large enough to leave its mark on the integrated straggling cross section.
\section{The stopping and straggling cross sections}
\begin{figure}
\includegraphics{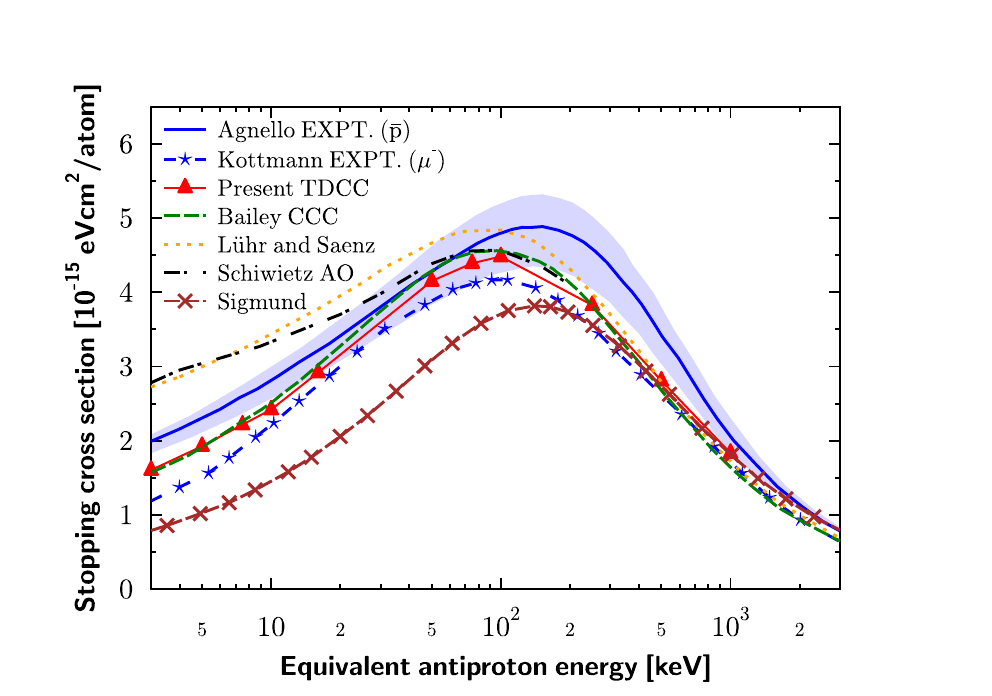}%
\caption{The stopping cross section $S(v_p)$ as a function of antiproton impact energy. The present TDCC results are compared to the experimental data for $\mathrm{\bar p}$  \cite{agnello,lodi04} (experimental uncertainty as shaded blue area), and $\mathrm{\mu^-}$ colliding with He~\cite{kottmann}, and  to other theoretical calculations: convergent close-coupling (CCC) of Bailey {\it et al.} \cite{bailey}; atomic-orbital close coupling of Schiwietz {\it et al.} \cite{schiwietz96}; semiclassical B-spline close-coupling calculations of L\"uhr {\it et al.} \cite{luhr}; and the binary collision theory of Sigmund {\it et al.} \cite{sigmund2001}. The TDCC results also include the contribution from nuclear stopping.
 \label{fig:compSexp}}
\end{figure}

The total stopping and straggling cross sections are calculated by integration over all impact parameters [Eqs. (\ref{eq:sint}) and (\ref{eq:tint})]. For the stopping cross section we can compare our present TDCC results for $\rm{\bar p}$ in He with available $\mathrm{\bar p}$ experimental data of Agnello {\it et al.} as reevaluated by Rizzini {\it et al.}  \cite{agnello,lodi04} and with experimental data of Kottmann for stopping of negatively charged muons ($\mu^-$) in He~\cite{kottmann}. Since the mass of $\mu^-$ ($=207$~a.u.) is large compared to that of the electron ($m_\mu \gg m_e$), inelastic electronic processes induced by isotachic (equal velocity) $\mu^-$ and $\mathrm{\bar p}$ projectiles should closely resemble each other and allow for a direct comparison of their stopping cross section. We also compare with other available theoretical results (Fig. \ref{fig:compSexp}).
Among those, the most advanced available approach is that  of Bailey {\it et al.} \cite{bailey} based on the convergent close-coupling (CCC) method in which the two-electron wave function is represented in a basis of target pseudostates and propagated in time numerically. Single ionization (SI0) and single excitation (EX0) are numerically accurately represented while double ionization and excitation ionization are approximated by a sequential independent event model.

With the exception of high projectile energies ($> 100$~keV), the agreement between the CCC and TDCC calculations is quite good, primarily because the dominant SI0 channel is treated in both approaches equivalently. The discrepancies observed for projectile energies above 100~keV can be attributed to the angular momentum basis truncation errors in the CCC calculations, in which the maximum angular momentum value  was $L_{max}=6$. This suppresses the formation of the high-energy part of the ionization spectrum (see Fig.\ref{fig:angdep}) and leads to the underestimation of the stopping cross section.

Remarkably, both state-of-the art calculations disagree with the $\mathrm{\bar p}$ experimental data by Agnello {\it et al.} \cite{agnello} as reevaluated in \cite{lodi04}. At low antiproton energies both the TDCC and CCC results lie outside the error bars.  Below $10$~keV, the contribution from nuclear stopping sets in. We have therefore also included these corrections. However, the result still lies outside the quoted error interval of the experiment (Fig.\ref{fig:compSexp}).  Most significantly, the stopping power maximum appears to be displaced in the experiment to higher collision energies (close to $150$~keV). As discussed in \cite{bailey} these discrepancies may result
in part from the complex processing of the experimental data which give only indirectly access to $S(v_p)$. Closer agreement is found with the experimental $\mu^-$ data, in particular the projectile velocity (or equivalent energy) for which the stopping cross section reaches its maximum coincides with that in the simulation. Yet, noticeable discrepancies in magnitude appear as well whose significance is difficult to assess in view of the unknown experimental uncertainties. 

Earlier calculations have been performed within the framework of one-electron models. They include the atomic-orbital close-coupling model of Schiwietz {\it et al.}~\cite{schiwietz96}, the electron-nuclear dynamics model by Cabrera-Trujillo {\it et al.}~\cite{cabrera05}, and the pseudostate close-coupling approach by L\"uhr {\it et al.} ~\cite{luhr}.  Contributions of two-electron processes to the stopping cross section are approximately included in these calculations employing an IEM. The agreement between these one-electron models \cite{schiwietz96,luhr} and the present two-electron calculations is good at high antiproton energies ($>200$~keV), while at lower antiproton energies the one-electron calculations overestimate the stopping cross section. It is of conceptual interest to identify the origin of this discrepancy. To this end, we have decomposed the TDCC results for $S(v_p)$ into the contributions due to the one-electron processes SI0 and EX0 and the two-electron processes DI and SI-EX. At intermediate energies 50~keV~$< E <$~200~keV the SI0 and EX0 contributions agree very well with the present SAE model and also with that of L\"uhr {\it et al.} \cite{luhr}. In this energy regime the discrepancy is thus due to the overestimation of uncorrelated multi-electron transitions within the IEM.
At even lower energies ($< 50$ keV) additional discrepancies appear already in the SI0 and EX0 contributions to stopping indicating that in this strongly non-perturbative regime electron correlation effects play an important role already in one-electron transitions. 

The binary collision theory for stopping by Sigmund et al.~\cite{sigmund2001} originally designed for swift heavy ions is based on an interpolation of the stopping numbers $L$ between the classical one-electron binary collision model at low collision speeds [Eq.(\ref{eq:L_Bohr})] and the Bethe limit at high speeds [Eq.(\ref{eq:L_Bethe})]. Multi-electron effects are indirectly included via shell corrections  and screening. Its application to $\rm{\bar p + He}$ yields qualitative agreement with the ab-initio calculations while systematically underestimating the stopping cross section below and around the stopping maximum. At high collision energies, the binary collision theory converges, by construction, to the Bethe limit and agrees quite well with the present TDCC results. For a critical comparison with the experiment it is worthwhile recalling the limitations of the present TDCC approach to stopping.  Deviations from a classical straight-line trajectory or diffractive scattering are neglected from the outset. Such effects are, however, expected to be negligible for equivalent projectile energies above a few keV (for $\mu^-$ slightly higher than for $\mathrm{\bar p}$). Furthermore, the present asymmetric angular momentum basis limiting the accessible angular momenta of the spectator electron of the primary collision event with the projectile to $l\le 1$ cannot reliably account for secondary violent electron-electron scattering events that have been identified in the equal energy sharing region of double ionization  by high-energy photons \cite{amusia,schoffler} and in the electronic Thomas scattering \cite{thomas,mergel97,fischer,Gudmundsson} mediating simultaneous charge transfer and ionization in charged particle collisions. Such processes have, however, negligible cross section compared to single ionization or excitation-ionization and are not expected to significantly influence the stopping cross section. Short-ranged non-electromagnetic interactions, including weak interactions are negligible as well since stopping is strongly dominated by distant collisions and long-range interactions. The main limitation is thus the non-relativistic treatment.  While for the numerical results  presented with $\mathrm{\bar p}$ energies up to 1~MeV corresponding to $\gamma\le 1.005$ relativistic corrections are still very small, at higher energies they may become significant. The high-energy behavior of stopping and straggling discussed here refers to the non-relativistic limit only.    

Despite its importance for characterizing the DET distributions, experimental results on straggling cross sections for gas targets are  still remarkably sparse \cite{bonderup,andersen1978,besenbacher1980,hvelplund75} as most of the measurements are performed for solid targets \cite{sigmundbook}. In particular, for $\rm \bar p$ on He neither experimental data nor numerical simulations appear to be available. The only available measurements somewhat related to the present calculations are those of Bonderup \emph{et al.}~\cite{bonderup}, and of Besenbacher et al.~\cite{besen1981}  performed for proton projectiles on a He gas target.

\begin{figure}
\includegraphics{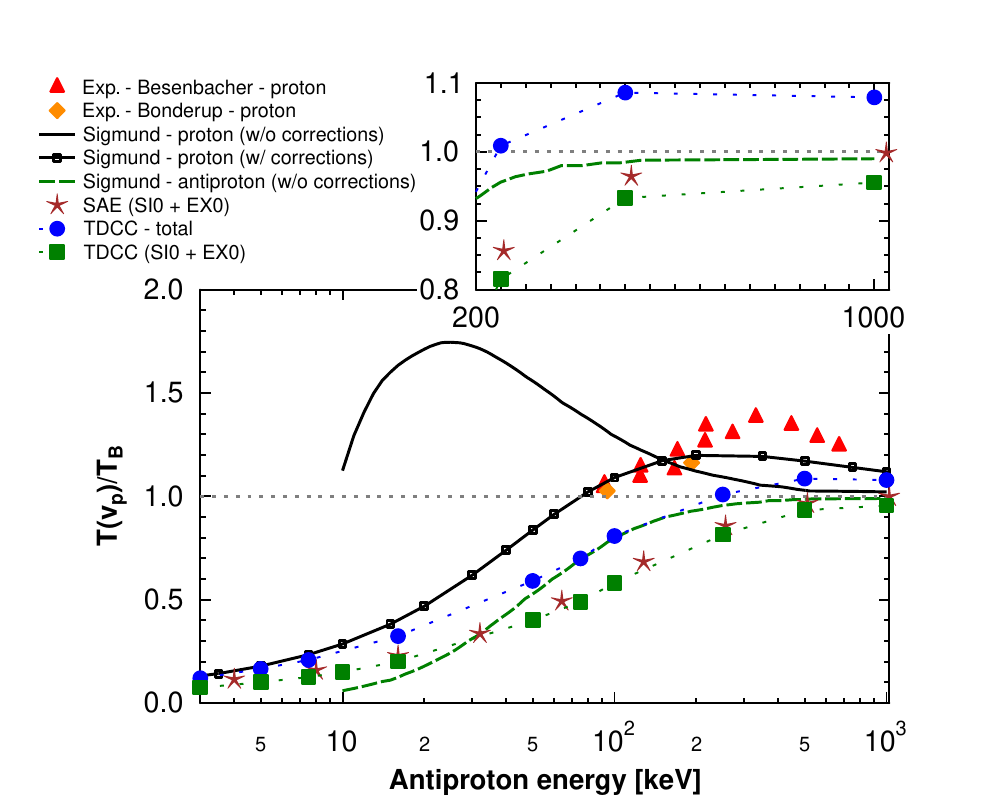}%
\caption{Comparison between straggling cross sections for protons and antiprotons colliding with helium, normalized to the Bohr straggling number $T_B=4\pi Z_p^2Z_T$. Shown are the present TDCC and SAE results for antiprotons, the experimental data by Bonderup {\it et al.} \cite{bonderup} and by Besenbacher et al.~\cite{besen1981} for protons. The analytic predictions by Sigmund for antiprotons and for protons are also shown, the latter both with ~\cite{sigmund2003,sigmund2010} and without~\cite{sigmundbook} corrections (see text). The smaller frame is a zoom-in on the high antiproton energy region showing the (non)convergence of the presented results towards the classical Bohr limit.
\label{fig:stragglingbohr}}
\end{figure} 

We present here first \emph{ab initio} straggling simulations for antiprotons using the fully correlated TDCC approach as well as the SAE model (Fig. \ref{fig:stragglingbohr}). The energy independent Bohr straggling cross section $T_B$ (Eq.~\ref{eq:TB}), based on the energy transfer in classical binary collisions with quasi-free electrons, gives the natural scale for straggling and provides a useful order-of magnitude estimate. We therefore display the experimental results for $\mathrm{p}$ on He and the theoretical predictions for $\mathrm{ \bar p}$ on He in units of $T_B$. Of particular interest is the convergence behavior of $T(v_p)$ towards $T_B$ at large collision energies as frequently assumed or implied. Indeed, the present SAE simulations as well as the TDCC restricted to one-electron processes, i.e. the sum of the SI0 and EX0 contributions agree very well with each other over the entire range of energies investigated (3~keV~$\le E \le$~1~MeV) and monotonically approach the Bohr limit $T_B$.

The analytic theory by Sigmund~\cite{sigmundbook} also predicts a monotonic increase towards $T_B$ for $\mathrm{\bar p}$ while for the charge conjugate projectile $\mathrm{p}$ this limit is approached from above and displays a peak around 200~keV. The peak is significantly reduced and moved to higher projectile energies when including the effect of multiple Bohr oscillators, and shell and screening corrections in the binary theory formalism~\cite{sigmund2003,sigmund2010}. The enhanced straggling for $\mathrm{p}$ originates from the combined effects of the Barkas contribution ($\sim Z_p^3$) \cite{barkas63} and the charge transfer channel absent for $\mathrm{\bar p}$.

\begin{figure}
 \includegraphics{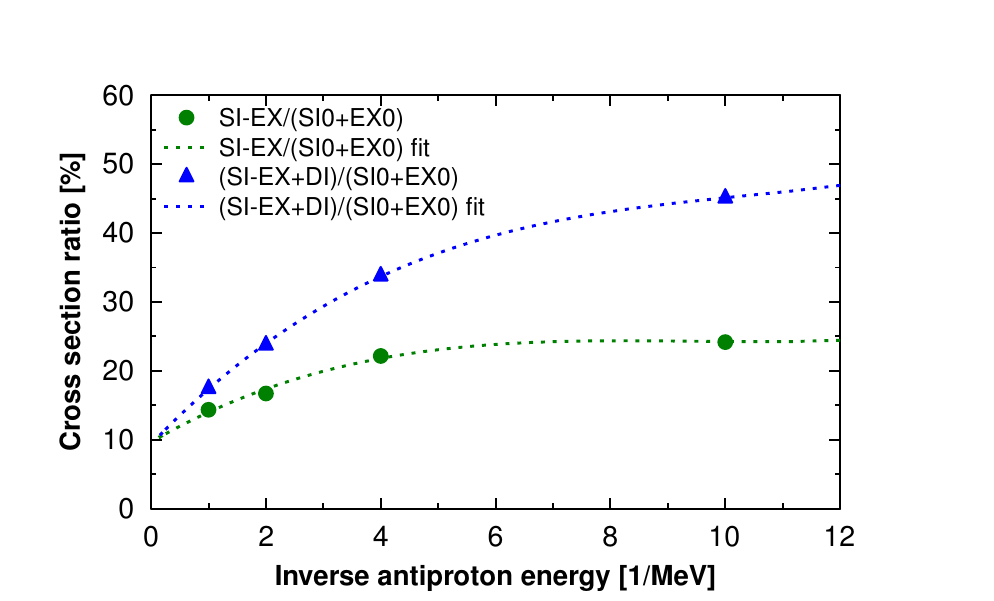}%
 \caption{ Straggling cross section ratios as a function of the inverse projectile energy. The TDCC data for the ratios $R$ are fitted to the following function: $R(E^{-1}) = R_0+a/E+b/E^2+c/E^3$. The $R_0$ asymptotic values of the ratios are 9.4\% [SI-EX/(SI0+EX0)] and 9.7\% [(SI-EX+DI)/(SI0+EX0)].    \label{fig:straggratio}}
 \end{figure}

The full TDCC, however, which includes the many-electron transitions does not  appear to converge to $T_B$ (see zoom-in in Fig.\ref{fig:stragglingbohr}), $T(v_p)$ lies about 10\% above $T_B$. The slight decrease by about 1\% of the TDCC straggling between 500~keV and 1~MeV antiproton energies (see zoom-in of Fig.\ref{fig:stragglingbohr}) might suggest a possible delayed convergence towards $T_B$, however, from above rather than from below. Based on both numerical evidence and analytic results for the non-relativistic high-energy limit this can be excluded since correlated two-electron processes, most importantly, single ionization accompanied by shake-up to excited states of He$^+$ (SI-EX) provide a finite contribution to $T(v_p)$ finite even as $E\rightarrow\infty$. It should be noted that a direct converged numerical calculation of $T(v_p)$ in the limit $v_p\rightarrow\infty$ is computationally not feasible within a given large, but finite-size angular momentum basis and FEDVR. Instead, we explore the asymptotic behavior of the two-electron processes to the total straggling cross section by fitting and extrapolating the ratios $R$ of the channels (SI-EX)/(SI0+EX0) and (SI-EX+DI)/(SI0+EX0) to the asymptotic expansion in powers of $E^{-1}$,
\begin{equation}\label{eq:ratio}
 R(E^{-1})=R_0+\frac{a}{E}+\frac{b}{E^2}+\frac{c}{E^3}.
\end{equation}
For both ratios the extrapolation yields nearly identical asymptotic limits of $R_0 \simeq 0.1$ (Fig. \ref{fig:straggratio}). While the DI channel provides a significant contribution at intermediate energies, the asymptotic behavior is dominated by the correlated shake-up.  
 It is the rapid decrease of the DI contribution between 500~keV and 1~MeV (or between 2 and 1~MeV$^{-1}$, Fig.~\ref{fig:straggratio}) which results in the slight decrease of $T(v_p)$ mentioned above (Fig.~\ref{fig:stragglingbohr}, inset). 
 
 The finite additional contribution of SI-EX and, to a lesser extent, of DI to the asymptotic straggling cross section beyond its one-electron limit is consistent with earlier analytic and numerical results of shake-up and shake-off processes in photoionization \cite{anderson93,tang97,aberg,dalgarno,byron,kabir}
which are closely intertwined with the analogous processes in charged-particle scattering \cite{burgdorfer97,mcquire}. Also in photoionization shake-up (i.e. SI-EX) and shake-off (DI) converge to a finite fraction of the SI0 cross section in the limit $E\rightarrow\infty$ with shake-up dominating over shake-off. The present findings are also consistent with earlier theoretical~\cite{lnagy99} and experimental~\cite{bailey95} data which show  the SI-EX/SI0 ionization cross section ratio to converge towards a constant nonzero value for large projectile velocities.

From the asymptotic behavior of $R(E^{-1})$ [Eq.\ref{eq:ratio} and Fig.\ref{fig:straggratio}] we estimate that the true (non-relativistic) high energy limit of straggling is $T\simeq 1.09T_B$ rather than $T_B$. Straggling is thus shown for the prototypical case of helium to be sensitive to multi-electron processes not accounted for by the Bohr model. This effect is expected to be more pronounced for heavier multi-electron atoms with plethora of available shake-up as well as correlated multiple shake-up-shake-off channels.

\section{Concluding Remarks}
We have presented a fully ab-initio simulation of the electronic energy loss distribution for antiproton scattering at He for antiproton energies ranging from $3$~keV to $1$~MeV using the TDCC method \cite{borbely14}. The first moment of this distribution, referred to as stopping cross section, and the second moment, the straggling cross section, are compared with other theoretical predictions and experiment when available. We have addressed the well-known discrepancy between several theoretical predictions~\cite{bailey,schiwietz96,luhr} and experimental data for the stopping cross section for $\mathrm{\bar p}$~\cite{agnello,lodi04} and $\mathrm{\mu^-}$~\cite{kottmann}. While we find slightly improved agreement with the $\mathrm{\bar p}$ experiment at high energies well above the stopping power maximum, the discrepancies to the data persist at lower energies while our TDCC results are in good accord with the recent CCC calculation \cite{bailey} both of which explicitly include electron correlation effects.  While all numerical simulations employing either an effective one-electron or the full two-electron time-dependent Schr\"odinger equation agree with each other on the projectile velocity of the stopping maximum, the stopping power maximum of the $\mathrm{\bar p}$ experimental data differ from these predictions. 
Compared to the $\mathrm{\bar p}$ data, better agreement is found between the $\mathrm{\mu^-}$ experimental data and the theoretical predictions, in particular on the position of the stopping maximum, however the magnitude of the $\mu^-$ stopping cross section is somewhat lower than the theoretical prediction for all equivalent energies. The considerable spread  and uncertainties in the
available experimental data suggests that further experimental tests are desirable.

Both the stopping cross section and the straggling cross section are shown to be influenced by electron correlation effects. In particular, the first \emph{ab initio} simulation for straggling reveals the importance of correlated multi-electron transitions. Ionization accompanied by excitation of the second electron provides a non-vanishing contribution even at high collision energies. This shake-up process is at the origin why the Bohr straggling number is not approached at high energies but surpassed. The present results provide the first benchmark data for the role of correlations in stopping and straggling for the simplest multi-electron system, helium, for which a full \emph{ab initio} description is still feasible. We expect such multi-electron transitions in heavier atoms and more complex targets to be of even greater importance.

\section{Acknowledgments}
The present work was supported by FWF-SFB049 (Nextlite), FWF-SFB041 (VICOM), doctoral college FWF-W1243(Solids4Function), WWTF MA14-002, by the National Research, Development and Innovation Office (NKFIH) Grant No. KH 126886, and by the high performance computing resources of the Babe\c{s}-Bolyai University. JF acknowledges funding from the European Research Council under grant ERC-2016-STG-714870 and by the Ministerio de Econom{\'\i}a y Competitividad (Spain) through a Ram\'on y Cajal grant. XMT was supported by a Grants-in-Aid for Scientific Research (JP16K05495) from the Japan Society for the Promotion of Science. Part of the calculation  was performed using COMA and Oakforest-Pacs supercomputers at the Center for Computational Sciences, University of Tsukuba.
\bibliography{stopping_ref} 
 
\end{document}